\documentclass[aps, prb, 10pt, twocolumn, nofootinbib, notitlepage, superscriptaddress]{revtex4-2}
\usepackage{natbib}
\usepackage{graphicx}
\usepackage{dcolumn} 
\usepackage{bm}
\usepackage{mathtools}
\usepackage{amssymb}
\usepackage{enumitem}
\usepackage[svgnames,dvipsnames]{xcolor}
\usepackage[normalem]{ulem}
\usepackage[mathlines]{lineno}
\usepackage[titletoc, title]{appendix}
\usepackage{bbold}
\usepackage{comment}
\usepackage{hyperref}
\usepackage{braket}

\newcommand{\beq}{\begin{equation}}
\newcommand{\eeq}{\end{equation}}
\newcommand{\beqa}{\begin{eqnarray}}
\newcommand{\eeqa}{\end{eqnarray}}

\begin{document}

\title{Quantum corrections to the Josephson dynamics: \\ a population-imbalance approach}

\author{Oliver Hideg}
\affiliation{Dipartimento di Fisica e Astronomia ``Galileo Galilei'', Universit\`a di Padova, Italy} 
\affiliation{Physikalisches Institut, Universit\"at Bonn, Germany} 
\author{Sofia Salvatore}
\affiliation{Dipartimento di Fisica e Astronomia ``Galileo Galilei'', Universit\`a di Padova, Italy}
\author{Luca Salasnich}
\affiliation{Dipartimento di Fisica e Astronomia ``Galileo Galilei'', Universit\`a di Padova, Italy} 
\affiliation{INFN Sezione di Padova, Padova, Italy} 
\affiliation{Padua QTech Center, Universit\`a di Padova, Italy}

\begin{abstract}
We investigate quantum corrections to the Josephson dynamics 
of two weakly coupled Bose-Einstein condensates using the 
population imbalance as the sole dynamical variable. 
Starting from the two-variable action, we derive the 
imbalance-only Lagrangian with a position-dependent mass 
and quantize it via symmetric operator ordering. The leading 
quantum corrections to the classical potential and mass are 
computed via the one-loop quantum effective action, using a 
covariant background-field method that fully accounts for 
the coordinate dependence of the mass. This yields explicit 
expressions for the effective potential and the effective 
mass, from which we derive the quantum-corrected Josephson 
frequency. Numerical comparison with exact diagonalization 
of the two-site Bose-Hubbard model shows that the 
imbalance-only formulation outperforms the complementary 
phase-only approach in the regime of weak interactions, 
which is the natural domain of validity of the 
population-imbalance description.
\end{abstract}

\maketitle

\section{Introduction}

The Josephson effect, that is the coherent tunneling of particles 
between two weakly coupled quantum systems, is one of the 
most remarkable manifestations of macroscopic quantum 
coherence. Originally predicted for superconductors 
\cite{Josephson}, it has since been observed in a variety 
of systems, including superfluid helium and, most relevantly 
for the present work, double-well trapped Bose-Einstein 
condensates \cite{Albiez, Spagnolli}. In the latter context, 
the Josephson dynamics is governed by two collective 
variables: the relative phase $\phi$ between the two 
condensates and the population imbalance $z = (N_1-N_2)/N$, 
whose mean-field equations of motion were derived by 
Smerzi and collaborators \cite{Smerzi}.

Beyond mean-field, quantum fluctuations modify the 
Josephson dynamics in a systematic and physically 
transparent way. A natural framework for computing 
these corrections is the quantum effective action 
\cite{coleman1973}, which organizes quantum effects 
as a loop expansion around the classical trajectory. 
This approach was applied to the bosonic Josephson 
junction in Ref.~\cite{furutani2022}, where the 
leading quantum correction to the Josephson frequency 
was derived in the limit of large interaction strength 
$UN\gg J$. More recently, Ref.~\cite{vianello2025} 
extended this treatment to arbitrary interaction 
strengths by deriving the one-loop quantum effective 
action for the relative phase $\phi$ alone, integrating 
out the population imbalance at the Gaussian level 
and using a covariant background-field method to 
account for the spatial dependence of the resulting 
effective mass. That work also identified a phase diagram 
delineating the regimes of validity of the $\phi$-only 
effective description. 

In this paper we develop the complementary formulation, 
in which the population imbalance $z$ is taken as the 
sole dynamical variable. This approach offers a 
different and independent perspective on the quantum 
corrections to the Josephson dynamics. Starting from 
the two-variable action $S[\phi,z]$, we derive the 
$z$-only Lagrangian and quantize it via symmetric 
operator ordering, obtaining a Schr\"odinger equation 
with a position-dependent mass. We then compute the 
leading quantum corrections to the classical dynamics 
using the one-loop quantum effective action with the 
covariant background-field method of 
Ref.~\cite{kleinert2002}, obtaining explicit expressions 
for both the effective potential $V_\mathrm{eff}(z)$ and 
the effective mass $m_\mathrm{eff}(z)$. As a consistency 
check, we verify that the correction to $V_\mathrm{eff}$ 
agrees with the result of the Ehrenfest theorem in the 
locally harmonic approximation. We then compute the 
quantum-corrected Josephson frequency and compare 
with exact diagonalization of the two-site Bose-Hubbard 
model, finding good agreement within the expected domain 
of validity of the $z$-only description.

\section{From the two-variable action to the z-only Lagrangian}

The starting point is the action for the Josephson dynamics 
of two weakly coupled Bose-Einstein condensates, expressed 
in terms of the relative phase $\phi(t)$ and the population 
imbalance $z(t) = (N_1(t)-N_2(t))/N$:
\beqa
S[\phi, z] &=& \int dt \Big[\frac{N\hbar z}{2}\dot\phi 
- \frac{UN^2}{4}z^2 
\nonumber 
\\
&+& JN\sqrt{1-z^2}\cos\phi\Big] \; .
\label{action}
\eeqa
Here $N$ is the total number of bosons, $J>0$ is the 
tunneling energy, and $U>0$ is the on-site interaction 
strength. The Euler-Lagrange equations derived from 
$S[\phi,z]$ are the mean-field equations of the 
Josephson dynamics \cite{Smerzi}:
\begin{subequations}
\label{eqmfs}
\begin{align}
\hbar\dot\phi &= \frac{2Jz}{\sqrt{1-z^2}}\cos\phi 
+ UNz \; , \label{eqmf1} \\
\hbar\dot z &= -2J\sqrt{1-z^2}\sin\phi \; . \label{eqmf2}
\end{align}
\end{subequations}
The conserved energy associated with the action 
(\ref{action}) is
\beq
\label{energy}
E(\phi,z) = \frac{UN^2}{4}z^2 - JN\sqrt{1-z^2}\cos\phi \; .
\eeq

To obtain an effective description in terms of $z$ alone, 
we integrate out $\phi$ at the classical level by solving 
Eq.~(\ref{eqmf2}) for $\phi$ and substituting back into 
the action. A simpler and more transparent approach is 
to work directly at the level of the equations of motion. 
In the regime of small phase oscillations around 
$\phi = 0$, one can set $\cos\phi \simeq 1$ and 
$\sin\phi \simeq \phi$, so that Eq.~(\ref{eqmf1}) gives
\beq
\phi = \frac{\hbar\dot z}{2J\sqrt{1-z^2}} \; ,
\eeq
where we have used Eq.~(\ref{eqmf2}) to express 
$\dot\phi$ in terms of $z$. More precisely, combining 
Eqs.~(\ref{eqmf1}) and (\ref{eqmf2}) and eliminating 
$\phi$, one finds that $z(t)$ satisfies the 
Euler-Lagrange equation of the $z$-only Lagrangian
\beq
L = \frac{1}{2}\,m(z)\,\dot z^2 - V(z) \; ,
\label{lag}
\eeq
where the $z$-dependent mass and potential are
\beq
m(z) = \frac{N\hbar^2}{4J\sqrt{1-z^2}} \; ,
\eeq
\beq
V(z) = \frac{UN^2}{4}z^2 - JN\sqrt{1-z^2} \; .
\eeq
This Lagrangian can also be derived 
by integrating out the phase $\phi$ from the action (\ref{action}) 
at the Gaussian level, following an approach analogous 
to that of Ref.~\cite{vianello2025} for the 
complementary $\phi$-only formulation. 

\section{z-only Hamiltonian}

Let us introduce the conjugate momentum
\beq
p_z = {\partial L\over \partial {\dot z}} = m(z) \, {\dot z} \; . 
\eeq
The Hamiltonian is then
\beq
H = p_z {\dot z} - L = {p_z^2\over 2 \, m(z)} + V(z) \; .
\label{ham}
\eeq
The conjugate momentum $p_z$ can be promoted to a momentum operator
${\hat p_z}$ as follows
\beq
p_z \to {\hat p_z} = - i \hbar \, \partial_z \;  
\eeq
such that $[z,{\hat p_z}] = i \hbar$.
In this way also the Hamiltonian function (\ref{ham}) can be promoted
to a Hamiltonian operator ${\hat H}$. A possibility is the following
\beq
H \to {\hat H} = {\hat p_z} {1\over 2 m(z)} {\hat p_z} + V(z) \; . 
\eeq
The symmetric operator ordering is chosen in order to guarantee
Hermiticity of the Hamiltonian operator. Explicitly we have
\beq
{\hat H} = - {2J\over N}
\partial_z \sqrt{1-z^2} \partial_z + V(z) \: .
\label{qham}
\eeq
Notice that the Hamiltonian operator (\ref{qham}) can also be obtained 
from the $z$-only Lagrangian (\ref{lag}) considering the short-time
propagator of Feynman's path-integral formulation. This quite standard 
procedure is slightly more complicated due to the presence of
a $z$-dependent mass. Following
this procedure one obtains the Schr\"odinger equation
\beq
i\hbar\,\partial_t\psi(z,t) 
= {\hat H} \psi(z,t) \; .
\eeq
with ${\hat H}$ given by Eq.~(\ref{qham}) and $\psi(z,t)$ the complex
probability amplitude of finding the bosonic system with
imbalance $z$ at time $t$.

\section{Semiclassical corrections via Ehrenfest theorem}

We now derive the equation of motion for the expectation value 
$\langle z(t)\rangle$ using the Ehrenfest
theorem \cite{ehrenfest1927,sakurai2020},
including leading-order quantum corrections.

Working with the Hamiltonian (\ref{qham}), from the Heisenberg equation
of motion one obtains
\beq
\frac{d}{dt}\langle z \rangle = \left\langle \frac{\hat{p}_z}{m(z)}
\right\rangle \; ,
\eeq
\beq
\frac{d}{dt}\langle \hat{p}_z \rangle = -\langle V'(z) \rangle 
+ \left\langle \frac{\hat{p}_z^2}{2} \partial_z \frac{1}{m(z)}
\right\rangle \; .
\eeq
This system does not close, since $\langle f(z)\rangle \neq
f(\langle z\rangle)$ for a nonlinear function $f$.

We expand around the expectation value by writing 
$z = \langle z\rangle + \delta z$, with $\langle \delta z\rangle = 0$
and $\sigma^2 = \langle (\delta z)^2\rangle$. 
Expanding $V(z)$ and $m(z)$ in Taylor series around $\langle z \rangle$
and retaining terms up to order $\sigma^2$, one finds
\beq
\langle V'(z)\rangle = V'(\langle z\rangle) 
+ \frac{1}{2}V'''(\langle z\rangle)\,\sigma^2 \; .
\eeq
The equations of motion then close at order $\sigma^2$, giving
\beq
m(\langle z\rangle)\, {d^2\over dt^2}{\langle z\rangle} 
= -V'(\langle z\rangle) 
- \frac{1}{2}V'''(\langle z\rangle)\,\sigma^2 \; . 
\eeq

Near the equilibrium point $z_0$, defined by $V'(z_0)=0$, 
the system behaves as a harmonic oscillator with local frequency
\beq
\omega_0(z) = \sqrt{\frac{V''(z)}{m(z)}} \; .
\eeq
For the ground state of the local harmonic oscillator, 
the zero-point variance is
\beq
\sigma^2 = \frac{\hbar}{2\,m(z_0)\,\omega_0(z_0)} \; .
\eeq
Substituting into the equation of motion, one finds that 
the quantum correction can be absorbed into an effective potential:
\beq
V_\mathrm{eff}^{(\mathrm{loc})}(z) = V(z) + \frac{\hbar}{2}\,\omega_0(z) 
= V(z) + \frac{\hbar}{2}\sqrt{\frac{V''(z)}{m(z)}} \; .
\label{veffloc}
\eeq
The quantum correction $\frac{\hbar}{2}\omega_0(z)$ is the 
local zero-point energy of the harmonic oscillator approximating 
the potential well at each point $z$. This locally harmonic 
result will be recovered as the leading term of the full 
one-loop calculation in the next section.

\section{Quantum effective action and effective potential}

The semiclassical effective potential of the previous 
section can be obtained more systematically, and 
generalized, via the quantum effective action 
\cite{coleman1973}, using the covariant background-field 
method for systems with coordinate-dependent mass 
\cite{kleinert2002,cametti2000}.

We expand the action
\beq
S[z] = \int dt \left[ {1\over 2} m(z) {\dot z}^2 - V(z) \right] 
\eeq
around the background dynamical variable $Z(\tau)$, which in standard QFT is the quantum average of $z(\tau)$,  setting   
\beq 
z(\tau) = Z(\tau) + \eta(\tau)
\eeq
with $\eta(t)$ representing quantum fluctuations. Performing a change of variables 
$\tilde\eta = \sqrt{m(Z)}\,\eta$ to render the 
path-integral measure flat, the fluctuation operator 
at second order in $\tilde\eta$ takes the form 
$-\partial_t^2 + \omega^2(Z)$, where the 
covariant fluctuation frequency is \cite{vianello2025,kleinert2002}
\beq
\omega^2(Z) = \frac{1}{m(Z)}\left[V''(Z) 
- \gamma(Z)\,V'(Z)\right] \; ,
\label{omegadef}
\eeq
with the Christoffel-like symbol
\beq
\gamma(Z) = \frac{m'(Z)}{2m(Z)} 
= \frac{z}{2(1-z^2)} \; .
\eeq
Performing the derivative expansion of the one-loop 
effective action for slowly varying $Z(t)$, one obtains 
\cite{kleinert2002,cametti2000}
\beq
\Gamma[Z] = \int dt \left[
\frac{m_\mathrm{eff}(Z)}{2}\,\dot{Z}^2 
- V_\mathrm{eff}(Z)
\right] + O(\hbar^2) \; ,
\label{q eff action}
\eeq
where the one-loop corrected potential and mass are
\beq
V_\mathrm{eff}(Z) = V(Z) + \frac{\hbar}{2}\,\omega(Z)
\label{veff}
\eeq
\beq
m_\mathrm{eff}(Z) = m(Z) + \frac{\hbar}{32}\,
\frac{\left[D\omega^2(Z)\right]^2}{\omega^5(Z)}
\label{meff}
\eeq
with $D\omega^2$ the covariant derivative of $\omega^2$:
\beq
D\omega^2(Z) = \frac{UNJ}{\hbar^2}
\frac{(3Z^2-4)Z}{2(1-Z^2)^{3/2}} 
+ \frac{J^2}{\hbar^2}\frac{Z}{(1-Z^2)^2} \; .
\label{domega2}
\eeq
Substituting the explicit expressions for $m(z)$ and 
$V(z)$, the fluctuation frequency (\ref{omegadef}) reads
\beq
\omega^2(Z) = \frac{UNJ}{2\hbar^2}
\frac{2-3Z^2}{\sqrt{1-Z^2}}
+ \frac{J^2}{2\hbar^2}\left(1 + \frac{1}{1-Z^2}\right) \; ,
\label{omegafull}
\eeq
and the full effective potential is
\beqa
V_\mathrm{eff}(Z) &=& \frac{UN^2}{4}Z^2 - \frac{JN}{2}\sqrt{1-Z^2}
\nonumber
\\
&+& \frac{1}{2}\Big[
\frac{UNJ}{2}
\frac{2-3Z^2}{\sqrt{1-Z^2}}
\\
&+& \frac{J^2}{2}\left(1 + \frac{1}{1-Z^2}\right)
\Big]^{1/2} \; .
\nonumber 
\label{veffexpl}
\eeqa
We note that in the locally harmonic limit (i.e.\ neglecting 
the $\gamma V'$ term in (\ref{omegadef})), the frequency 
reduces to $\omega_0(z) = \sqrt{V''(z)/m(z)}$ and one 
recovers the Ehrenfest result (\ref{veffloc}). 
The covariant treatment introduces a correction proportional 
to $\gamma(Z) V'(Z)$ which vanishes at $Z=0$ but is 
important for oscillations around symmetry-breaking 
minima at $Z \neq 0$.

\section{Quantum correction to the Josephson frequency}
\label{sec:quantum eff action}
The quantum-corrected Josephson frequency is obtained 
by expanding $V_\mathrm{eff}(Z)$ and $m_\mathrm{eff}(Z)$ 
around $Z=0$ and evaluating the small-oscillation frequency
\beq
\Omega_J = \sqrt{\frac{V''_\mathrm{eff}(0)}{m_\mathrm{eff}(0)}} \; .
\eeq
From Eq.~(\ref{omegafull}), the second derivative of 
$V_\mathrm{eff}$ at $Z=0$ receives a contribution from 
the quantum correction $\frac{\hbar}{2}\omega(Z)$, while 
the effective mass $m_\mathrm{eff}(0)$ also differs from 
the classical mass $m(0)$ through the correction 
(\ref{meff}). Combining both contributions, one finds
\beqa
\Omega_J^{(z)} &=&  \omega_J \sqrt{ 1- \frac{1}{2N}\frac{2\Lambda^2 +\Lambda - 1}{(\Lambda+1)^{5/2}}}
\nonumber
\\
&=&\omega_J \sqrt{ 1- \frac{1}{2N}\frac{2\Lambda-1}{(\Lambda+1)^{3/2}}} \; ,
\label{omegacorr}
\eeqa
where $\Lambda = UN/2J$ is the dimensionless interaction 
parameter and $\omega_J = \sqrt{2J(UN+2J)}/\hbar$ 
is the classical Josephson frequency.
 
It is instructive to compare this result with that 
obtained in the $\phi$-only formulation of 
Ref.~\cite{vianello2025}, where the corrected frequency reads
\beq
\Omega_J^{(\phi)} = \omega_J\sqrt{1- \frac{1}{2N}\frac{\Lambda+3}{(\Lambda+1)^{1/2}}} \; .
\label{omegaphi}
\eeq
The two results have the same structure but differ 
at finite $\Lambda$, reflecting the different 
approximations involved in integrating out one of 
the two degrees of freedom in each formulation.
\begin{comment}
In the limit $\Lambda = UN/2J\gg 1$, both expressions 
(\ref{omegacorr}) and (\ref{omegaphi}) reduce to the 
same universal result, independent of the choice of 
variable:
\beq
\Omega_J = \omega_J\left(1 - \frac{1}{4N}
\sqrt{\frac{U}{2JN}}\right) \; ,
\label{universal}
\eeq
which was first derived in Ref.~\cite{furutani2022} 
in the approximation of constant mass. The agreement 
of the two formulations in this limit provides a 
strong consistency check on both results.
\end{comment}
\section{Numerical Comparison with Exact Diagonalization}
\label{sec:numerical}
We assess the validity of the quantum correction to the frequency of small-amplitude oscillations around the Josephson fixed point by comparing the mean-field $\hbar \Omega_J$ to the energy gap between the ground state and first excited state $E_1-E_0$ of the Hamiltonian operator ${\hat H}$ of the two-site  
Bose-Hubbard model:
\begin{equation}
    \hat H = -J\left(\hat a_1^\dagger \hat a_2 + \hat a_2^\dagger \hat a_1\right) + \frac{J\Lambda}{N} \sum_{i=1,2} \hat a_i^\dagger \hat a_i^\dagger \hat a_i \hat a_i 
    \label{2site BH ham}
\end{equation}
where ${\hat a}_i$ and ${\hat a}_i^{\dagger}$ are the ladder operators. The quantum formulation based on the Hamiltonian (\ref{2site BH ham}) is 
equivalent to the path integral approach based on the action functional of Eq. \eqref{action}, for details see Ref. \cite{vianello2025}.  
Furthermore, we contrast the mean-field \eqref{eqmfs}, and the quantum effective mean-field \eqref{q eff action} to the exact quantum dynamics described below.

The quantum evolution of the imbalance is calculated by diagonalizing the Hamiltonian $\hat H$ and writing the propagator in the energy eigenbasis as $e^{-i\hat H t/\hbar}\ket{E_j} = e^{-iE_j t/\hbar}\ket{E_j}$. Thus, given an initial state $\ket{\psi_0}$, the quantum evolution of the imbalance is given by:
\begin{equation}
    \braket{\psi_0|\hat z(t)|\psi_0} = \sum_{j,k=0}^N Z_{jk}e^{-i\omega_{jk}t}
    \label{z fourier}
\end{equation}
where the frequencies correspond to transitions between energy levels: $\omega_{jk} = (E_k - E_j)/\hbar$, and the amplitudes are
\begin{equation}
    Z_{jk} = \braket{\psi_0|E_j}\braket{E_j| \hat z|E_k}\braket{E_k|\psi_0}
\end{equation}
where $\hat z = (\hat a_1^\dagger \hat a_1 - \hat a_2^\dagger \hat a_2)/N$ is the fractional imbalance operator.
Similarly to Ref. \cite{vianello2025} we calculate the quantum evolution of the phase difference as:
\begin{equation}
    \braket{\hat \phi(t)} =- \mathrm{arg}\left(\braket{\psi_0|\hat a^\dagger_1 \hat a_2|\psi_0}\right)
\end{equation}
In order to describe small-amplitude oscillations around the Josephson fixed point, as in Ref. \cite{vianello2025}, we take the initial state $\ket{\psi_0}$ to be the atomic- or spin coherent state:
\beqa
    \ket{z_0,\phi_0} &=& \sum_{n=0}^N \binom{N}{n}^\frac{1}{2} \left(\frac{1+z_0}{2}\right)^{\frac{n}{2}} \left(\frac{1-z_0}{2}\right)^{\frac{N-n}{2}} 
    \nonumber 
    \\
    &\times& 
    e^{-in\phi_0}\; \ket{n}_1\ket{N-n}_2
\eeqa
with $z_0=0$ and $\phi_0 = 0.1$. These initial conditions lead to small amplitude oscillations. In the quantum case, the atomic coherent state with these "initial conditions" favours the states $\ket{E_0}$ and $\ket{E_1}$, and strongly suppresses the higher energy states, so that the oscillations in Eq. \eqref{z fourier} are dominated by the frequency equal to the gap $E_1-E_0$.

\begin{figure}
    \centering
    \includegraphics[width=\linewidth]{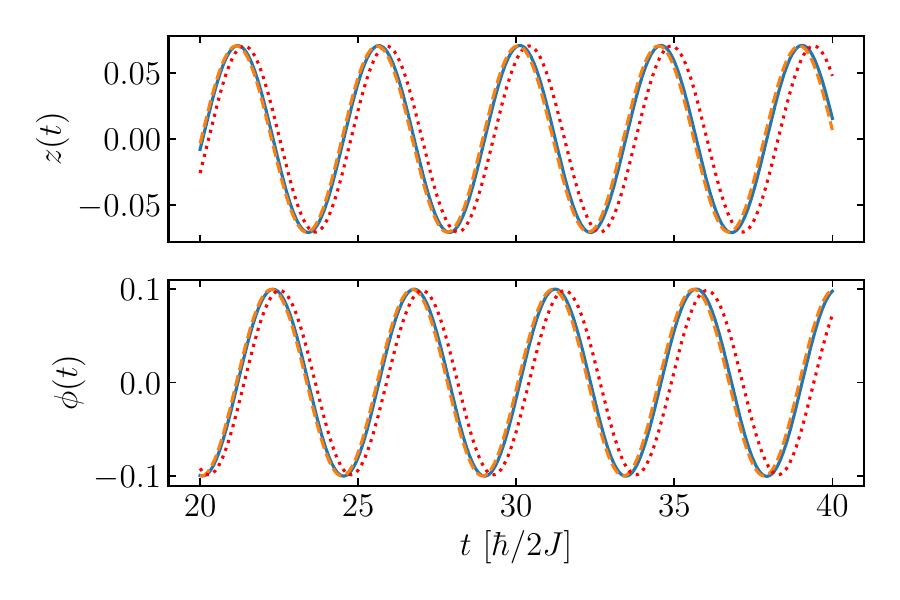}
    \caption{The exact evolution (solid, blue) 
    of $\braket{\hat z(t)}$ and $\braket{\hat \phi(t)}$, the mean-field solution (dotted, red) and the quantum effective mean-field (dashed, orange) for $\Lambda = 1$ and $N=40$. Initial conditions were set to $z_0=0$ and $\phi_0 = 0.1$. The quantum effective dynamics is more accurate than the mean-field.}
    \label{fig:jos_lambda}
\end{figure}

\begin{figure}
    \centering
    \includegraphics[width=\linewidth]{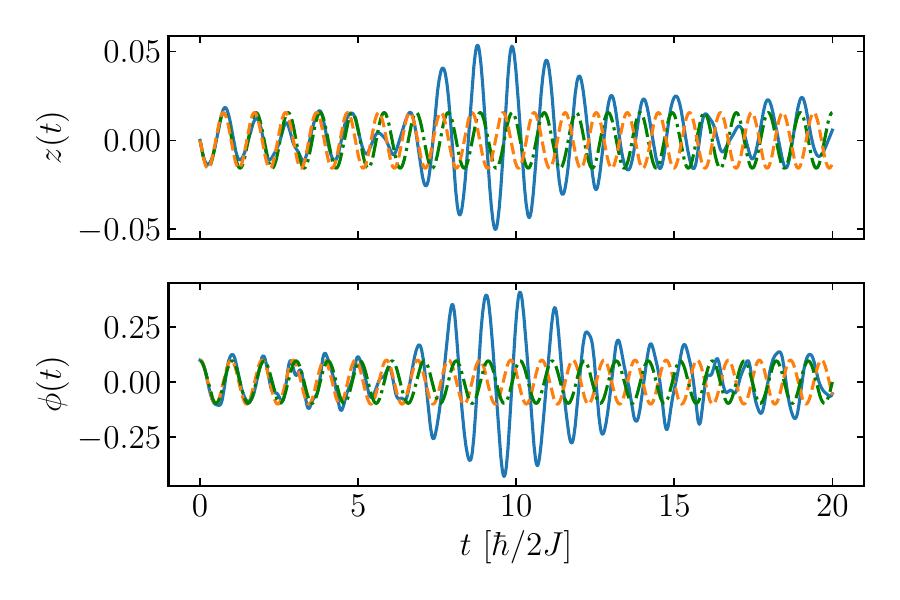}
    \caption{The exact evolution 
    (solid, blue) 
    of $\braket{\hat z(t)}$ and $\braket{\hat \phi(t)}$ and the quantum effective mean-field (dashed, orange) for $\Lambda = 40$ and $N = 40$. The initial conditions where set to $z_0 = 0$ and $\phi_0 = 0.1$. The mean-field is not shown as it almost completely overlaps with the quantum effective mean-field. Instead, we show a combination: the mean-field elliptic cosine with the only-$z$ corrected frequency \eqref{omegacorr} (dash-dot, green). The hybrid approach benefits from both the quantum correction and the elliptic correction to the time period \eqref{elliptic period}. However, neither can account for the amplitude modulation caused by higher energy levels influencing the dynamics in the fully quantum case.}
    \label{fig:fock_lambda}
\end{figure}

\begin{figure}
    \centering
    \includegraphics[width=\linewidth]{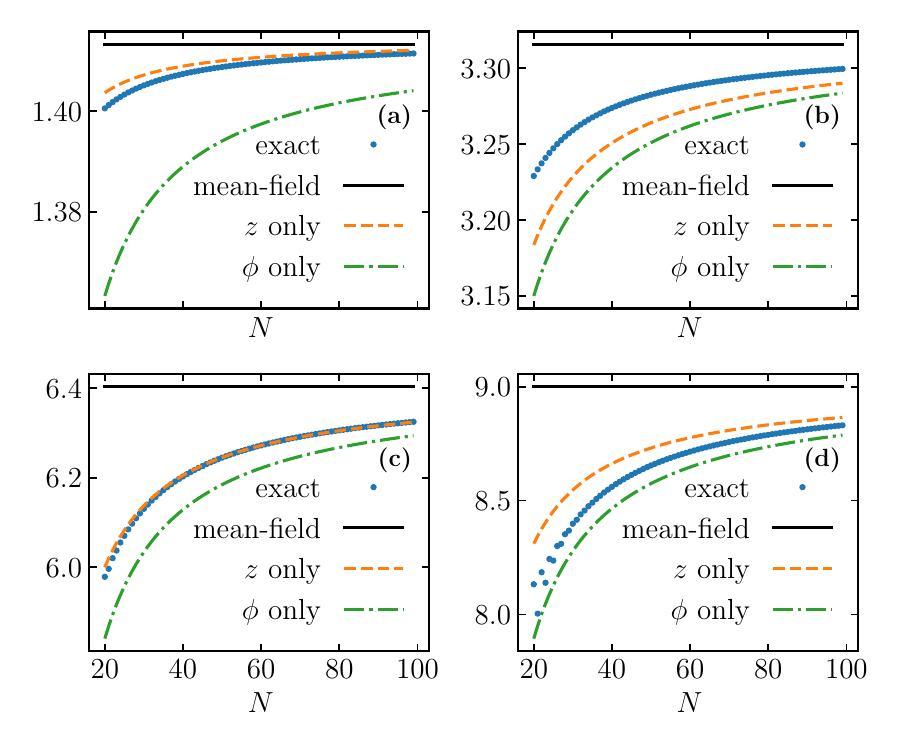}
    \caption{The four plots show the predicted Josephson oscillation frequency for $\Lambda = 1, 10, 40, 80$ together with the energy gap between the ground state and first excited state of the Hamiltonian obtained numerically as a function of the number of bosons $N$. To be able to compare them, the vertical axis shows energy in units of $2J$ and frequency in units of $2J/\hbar$. The imbalance only prediction (dashed) is closer to the numerical (dots) than the phase only prediction (dash-dot) for $\Lambda \lesssim 80$. The mean-field (solid), that is independent of $N$, is accurate only as $N \to \infty$.}
    \label{fig:gap freqs}
\end{figure}

The mean-field dynamics are described by equations of motion Eq. \eqref{eqmfs}. They are exactly solvable \cite{Smerzi}. In particular, the anharmonic oscillations around the Josephson fixed point $(z,\phi ) = (0,0)$ are analogous to the anharmonic oscillations of the simple pendulum around its stable fixed point; both given by the Jacobi elliptic cosine:
\begin{equation}
    z_\text{cl}(t) = z_+\mathrm{cn}(\omega t - \alpha_0,k).
    \label{elliptic cosine}
\end{equation}
We shall list the parameters that appear in this expression with a short explanation for each and show the limit for $z_0 \equiv 0$ and $\phi_0 \to 0$. A useful quantity is the dimensionless energy per particle:
\begin{equation}
    \mathcal E = \frac{E(\phi_0,z_0)}{JN} \to -\cos(\phi_0) \approx -1 + \phi_0^2
\end{equation}
where $E(\phi,z)$ is the one in Eq. \eqref{energy}.
The initial conditions $z_0=0$ and $|\phi_0|\ll 1$ ensure that the (classical) energy is close to the absolute minimum, the condition of Josephson oscillations.
\begin{equation}
    z_+^2 = \frac{2}{\Lambda^2}\left[(\Lambda \mathcal E-1) + \sqrt{\Lambda^2 +1-2\Lambda \mathcal E}\right] \to \frac{2\phi_0^2}{\Lambda(\Lambda + 1)}
\end{equation}
The amplitude of Josephson oscillation in $z$ is smaller than that in $\phi$.
\begin{equation}
    k^2 = \frac{1}{2}\left[1+\frac{\Lambda \mathcal E-1}{\sqrt{\Lambda^2 + 1 - 2\Lambda \mathcal E}}\right] \to \frac{\phi_0^2\Lambda^2}{4(\Lambda+1)^2}
\end{equation}
where $k$ is the elliptic parameter which measures the anharmonicity of Josephson oscillation.
\begin{equation}
    \omega = \frac{\sqrt[4]{\Lambda^2 + 1 - 2\Lambda \mathcal E}}{\hbar / 2J}\to \frac{\sqrt{\Lambda +1}}{\hbar/2J}\left[1-\frac{\phi_0^2\Lambda}{4(\Lambda + 1)}\right]
\end{equation}
where $\omega$ is the elliptic frequency. The period of oscillation is given by
\begin{equation}
    T = \frac{4K(k)}{\omega} \to \frac{2\pi}{\omega} \left[1+\frac{\phi_0^2 \Lambda^2}{16(\Lambda+1)^2}\right].
    \label{elliptic period}
\end{equation}
This demonstrates well how much the elliptic time period differs from the harmonic one as a function of the initial displacement.
Lastly, the initial phase is given by
\begin{equation}
    \alpha_0 = 3F\left(\arccos{\frac{z_0}{z_+}}\,,k\right) \to 3K(k)
\end{equation}
where $F(\varphi,k)$ and $K(k)$ are the incomplete and complete elliptic integral of the first kind. This is analogous to the harmonic approximation of the simple pendulum for small amplitude oscillations.
Plots of the exact quantum and mean-field dynamics can be found in Figs. \ref{fig:jos_lambda} and \ref{fig:fock_lambda}. In the region of parameter space where the mean-field solution is a good description of the quantum dynamics ($z,\phi \ll 1$), the $z$-only quantum effective mean-field does improve the standard mean-field (Fig. \ref{fig:jos_lambda}). Whereas, in the only phase approach, the anharmonicity was too large for the quantum effective mean-field to give a meaningful improvement for $\Lambda \lesssim 10$ \cite{vianello2025}. Furthermore, in this regime, the only-phase gives a poorer estimate for the gap of the quantum Hamiltonian \eqref{2site BH ham} (Fig. \ref{fig:gap freqs}) For $10\lesssim \Lambda\lesssim80$, the exact quantum dynamics deviates from the quantum effective- and standard mean-field during the first couple of oscillations. The quantum effective- and standard mean-field are almost identical (Fig. \ref{fig:fock_lambda}). The only-$z$ Josephson frequency \eqref{omegacorr} provides a better approximation for the gap of the Hamiltonian operator \eqref{2site BH ham}. This can be supplemented by adding an elliptic correction from \eqref{elliptic period}, by simply inserting the quantum corrected frequency $\Omega_J^{(z)}$ into the elliptic cosine of \eqref{elliptic cosine}. This combined approach is plotted too in Fig. \ref{fig:fock_lambda}, it is better at estimating the dominant frequency in the longer term. However, it cannot account for the amplitude modulation that is caused by higher energy modes influencing the dynamics in the fully quantum treatment. These modulations become more prevalent for larger values of $\Lambda$.

To compare how accurately the approach using the only-$z$ and only-phase quantum effective action can estimate the true gap of the quantum Hamiltonian \eqref{2site BH ham}, we plotted their predictions in Fig. \ref{fig:gap freqs} together with the standard mean-field's prediction for the Josephson dynamics for $\Lambda = 1,10,40,80$. It shows that for lower values of $\Lambda$ the only-$z$ approach is much closer to the numerical value of the gap. And, around the value $\Lambda \sim 80$, the only-phase approach becomes more accurate. We ascribe this to the difference in approximating either $z$ or $\phi$ to the quadratic order in \eqref{action} to evaluate its integral at the Gaussian level.


\section{Conclusions}

In our work we have investigated quantum corrections to the dynamics of Josephson oscillations of the relative imbalance in a bosonic Josephson junction. We derived the effective imbalance-only action from the Lagrangian function of collective variables $(\phi,z)$ by a quadratic approximation in $\phi$. This is analogous to the treatment in \cite{vianello2025} where the approximation was done in $z$ instead. With the rules of canonical quantization, and symmetric ordering, we derive a quantized Hamiltonian and a corresponding Schrödinger equation. Utilizing the Ehrenfest theorem we arrive at a locally harmonic quantum correction to the potential. In Section \ref{sec:quantum eff action} we use the quantum effective action approach to obtain finite-size quantum corrections to the mass and potential. These modify the mean-field equations and provide a quantum correction to the Josephson frequency. In Section \ref{sec:numerical} we compare the standard- and quantum effective mean-field with the exact numerical dynamics obtained by diagonalizing the two-site Bose-Hubbard Hamiltonian and using atomic coherent states to interpret the classical phase space in the quantum formulation.
We find that the finite size correction to the Josephson oscillations features a trade-off between the valuable correction to the Josephson frequency and the disadvantage of the quadratic approximation in $\phi$ that loses information about the anharmonicity. We also see that, in the Fock-regime ($\Lambda \gtrsim N$) the amplitude modulations introduced by higher energy states cannot be explained in our analytic approach. We show a combined approach as a possibility to overcome the trade-off between anharmonicity and corrected frequency, by simply using the corrected frequency in the solution of the standard mean-field equations.
Future work could involve a quantum effective action of the standard mean-field dynamics that would truly overcome the trade-off described above. Furthermore it would allow for corrections to the inverted and self-trapped oscillations and allow for analysis of the dynamics at finite temperature.

{\bf Acknowledgements}. 
The authors thank Cesare Vianello for useful suggestions. 
This work is partially supported by the `Iniziativa Specifica Quantum' of
INFN, by the Project ``Frontiere Quantistiche'' (Dipartimenti di 
Eccellenza) of the Italian Ministry of University and Research
(MUR), by the European Union-Next Generation EU within the National
Center for HPC, Big Data and Quantum Computing
(Project No. CN00000013, CN1 Spoke 10: `Quantum
Computing'), and by the EU Project PASQuanS 2
`Programmable Atomic Large-Scale Quantum Simulation'.

\end{document}